# Observation of Topological Hall Effect in Synthetic Antiferromagnetic Skyrmion System

Xinbao Geng[1#], Guanqi Li[1#], Zhongxiang Zhang [2,3], Wenjing Hu[1], Wenjing Zhong[1], Xiaoming Xiong[1], Yongbing Xu[1,4,5], Zhendong Chen[1*], Junlin Wang[1,4*], Xiangyu Zheng[2,3*], Jing Wu[1,4*]

[1] *School of Integrated Circuits, Guangdong University of Technology, Guangzhou, 510006, China*

[2] *Fert Beijing Institute, MIIT Key Laboratory of Spintronics, School of Integrated Circuit Science and Engineering, Beihang University, Beijing, 100191, China*

[3] *China National Key Laboratory of Spintronics, Hangzhou International Innovation Institute, Beihang University, Hangzhou,311115, China*

[4] *York-Nanjing International Center for Spintronics (YNICS), School of Physics, Engineering and Technology, University of York, York YO10 5DD, UK*

[5] *National Key Laboratory of Spintronics, Nanjing University, Suzhou, 215163, China*

[#] Xinbao Geng and Guanqi Li contributed equally to this work.

[*] Corresponding authors: chenzd0105@gdut.edu.cn, junlin.wang@gdut.edu.cn, xz1867@buaa.edu.cn, jing.wu@gdut.edu.cn.

## Abstract

Synthetic antiferromagnetic skyrmions have emerged as promising candidates for next-generation high-speed and highly integrated spintronic devices, owing to their exceptional properties such as high driving velocity, nanoscale dimensions, and the absence of the skyrmion Hall effect. In this work, we report the experimental observation of the topological Hall effect in both compensated and non-compensated synthetic antiferromagnetic skyrmion systems based on $[Pt/Co/Ru]_2$ trilayer. The antiferromagnetic skyrmions are further demonstrated to be robust in these synthetic antiferromagnets under zero-field. Our first-principles calculations and transmission electron microscope measurements show that the observed topological Hall Effect is related to the skyrmion structures in the Ru and Pt layers induced by the magnetic proximity effect and atomic diffusion. This work examines the role of the magnetic proximity effect in the topological Hall effect of SAF skyrmions in these systems.

Magnetic skyrmions, as a topologically protected vortex-like spin texture, have garnered significant research interest in recent years due to their unique physical properties and potential technological applications[1–7]. These magnetic textures exhibit soliton-like characteristics[8–12], enabling their controlled motion, mutual interactions, and distinct dynamic behaviors when subjected to external excitation, particularly through electric current manipulation[13–15]. Such remarkable features position skyrmions as promising candidates for next-generation high-density, low-power-consumption memory devices and innovative computing architectures, such as artificial neural networks[16–18]. However, practical application of ferromagnetic skyrmions faces two fundamental challenges. Firstly, due to the inherent skyrmion Hall effect, the ferromagnetic skyrmions show undesirable transverse deflection during current-driven motion, leading to eventual annihilation at device boundaries[19–21]. Secondly, the dipolar interaction induces a limitation to the minimum achievable size of ferromagnetic skyrmions[22]. These two limitations severely limit the application of ferromagnetic skyrmions.

An alternative approach to overcome these issues is the realization of antiferromagnetic skyrmions in synthetic antiferromagnets (SAFs). In SAFs, their magnetization is coupled through Ruderman-Kittel-Kasuya-Yosida (RKKY) coupling, resulting in zero net magnetization[23–25]. This configuration enables the stabilization of skyrmions with significantly reduced dimensions due to the suppression of the dipolar interactions[26]. Compared with that in single-phase antiferromagnets, the Dzyaloshinsky-Moriya interaction (DMI) in SAFs can be conveniently modulated through engineering the heavy-metal/ferromagnet interface[27,28], which is essential for the robustness of skyrmions at room temperature. The compensation of the topological charge in SAF skyrmions can significantly suppress the skyrmion Hall effect, allowing for SAF skyrmion rectilinear motion along the applied current direction[29–31]. Recent studies also point out that SAF skyrmions can move at the velocity of up to 900 m/s along the current direction by driving them with current-induced spin-orbit torque (SOT)[26,31–38]. These advantages make SAF skyrmions ideal carriers of information in highly integrated, high speed, and low energy dissipation spintronic devices. The magnetic structure, nucleation, and motion of SAF skyrmions have been investigated by several experimental techniques[29,39,40]. However, there is a lack of electric transport characteristics of the SAF skyrmions and the possible existence of the topological Hall effect (THE) has not been revealed so far.

In this work, we report the observation of the THE in a $[Pt/Co/Ru]_2$ based SAF skyrmion system. Robust SAF skyrmions under zero and finite external magnetic fields were observed in both non-compensated and compensated SAF systems using magnetic force microscopy (MFM) and scanning nitrogen-vacancy center microscope (SNVM). Significantly, Hall effect measurements revealed the emergence of the non-zero THE even in the case of complete magnetic moment compensation. First-principles calculations and transmission electron microscope (TEM) measurements confirm that the net magnetic moments can exist in the Pt and Ru layers induced by the magnetic proximity effect and the interfacial atomic diffusion, leading to the formation of skyrmions in the Pt and Ru layers and thus the observed THE. These results suggest significant roles of the magnetic proximity effect and atomic diffusion in the SAF skyrmion system and provide a possible approach to achieve the electronic detection of SAF skyrmions.

It is well known that the THE in ferromagnetic skyrmion systems is related to the topologically non-trivial magnetic texture of skyrmions. The $\rho_{\mathrm{H}}^{\mathrm{T}}$ can be expressed by the following equation:

$$\rho_{\mathrm{H}}^{\mathrm{T}} = \frac{\rho_0^{\mathrm{T}}}{4\pi} \iint \mathrm{d}^2\boldsymbol{r}\boldsymbol{m} \cdot \left(\frac{\partial \boldsymbol{m}}{\partial x} \times \frac{\partial \boldsymbol{m}}{\partial y}\right) \quad \text{Eq. (1)}$$

where $\rho_0^{\mathrm{T}}$, $\boldsymbol{r}$, and $\mathrm{m}$ are the topological Hall coefficient, the spatial coordinate, and the normalized magnetic moment at position $\boldsymbol{r}$, respectively. The integral part is defined as the skyrmion number $N_{\mathrm{sk}}$, which describes the topological properties of skyrmions. Eq. (1) indicates that for ferromagnetic skyrmions, the magnitude of the THE is independent of the external magnetic field or the size of the skyrmions, and depends on $N_{\mathrm{sk}}$ and the skyrmion density. However, for a single compensated SAF skyrmion, the value of $N_{\mathrm{sk}}$ is zero due to its zero net moment, thus fully compensated SAF skyrmion systems should exhibit neither THE nor skyrmion Hall effects, as shown in Fig. 1(a). Nevertheless, magnetic proximity effects in multilayer systems may induce weak net moments in non-magnetic layers, which could lead to detectable THE signals [Fig. 1(b)]. The THE induced by the non-magnetic layers will be discussed in this work.

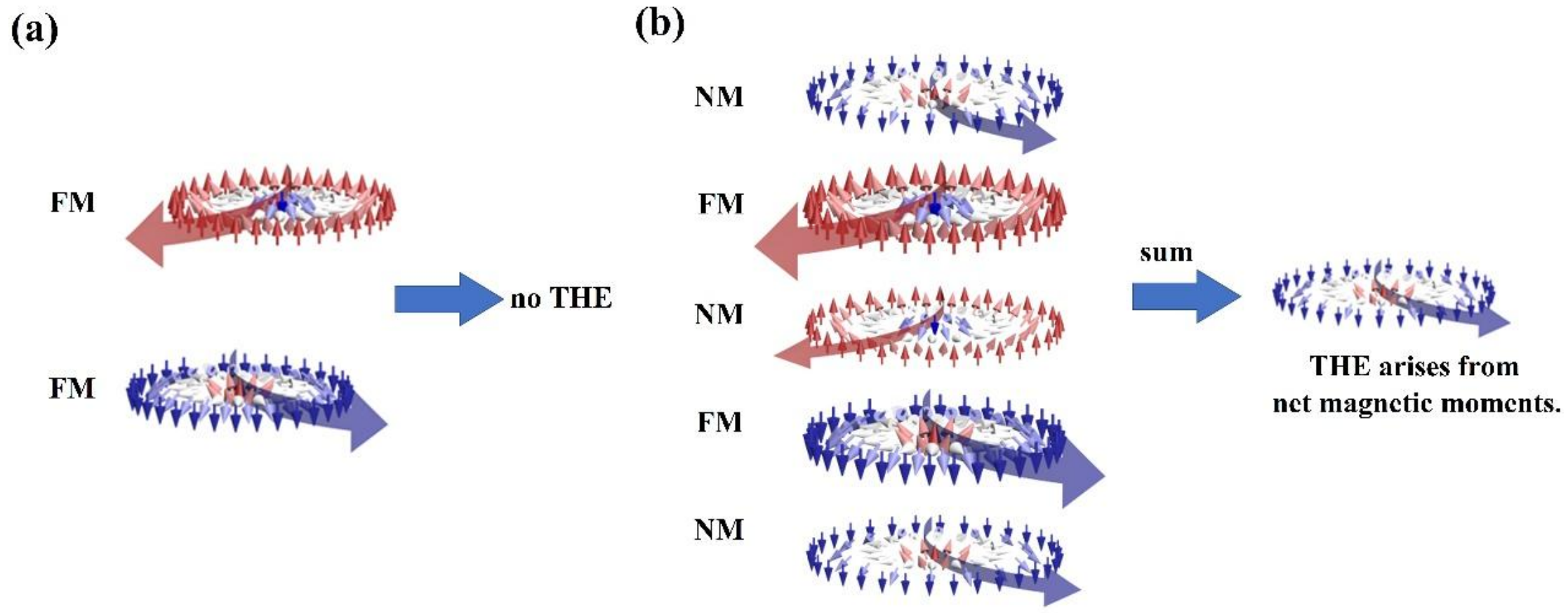

**FIG. 1.** Schematic illustration of the emergence of the THE in the compensated synthetic antiferromagnetic skyrmion system. (a) In the fully compensated SAF skyrmion system, THE signals from the top and bottom layers are equal in magnitude but opposite in direction, leading to a net cancellation. (b) Illustration of skyrmions in all the ferromagnetic (FM) and non-magnetic (NM) layers in the multilayer samples, as well as the skyrmion formed by the net magnetic moment. The arrow shows the electron trajectory induced by the skyrmion's emergent electromagnetic field.

In this work, two series of SAF samples based on $[Pt/Co/Ru]_2$ multilayers with different Co layer thicknesses were fabricated by magnetron sputtering. The first series possesses a non-compensated SAF structure: $Si/SiO_2$/[buffer layer]/Pt (0.6 nm)/Co ($x_1$)/Ru (0.5 nm)/Pt (0.6 nm)/Co ($x_2$)/Ru (0.5 nm)/Ta (5 nm), where both $x_1$ and $x_2$ varied from 1 nm to 1.8 nm, and $x_1 - x_2 = 0.2$ nm. The second series possess a fully compensated SAF structure with the same composition as the non-compensated SAF structure but with $x_1 = x_2$. Then atomic force microscope (AFM) and TEM measurements were performed to character the surface morphology and the structure of the samples. Detailed parameters for sample preparation, specific sample structures, and surface morphology can be found in Supplementary Materials Section 1 and Fig. S1. The optimization process of the sample parameters is provided in Supplementary Materials Section 2 and Fig. S2. The in-plane and out-of-plane hysteresis loops of the samples were measured via vibrating sample magnetometer (VSM) to confirm their magnetic anisotropy. Subsequently, MFM and SNVM tests were performed to obtain the images of the SAF skyrmions. During the MFM tests, different magnetic fields were applied along the film's normal direction. SNVM measurement parameters: NV-to-sample distance ~250 nm; NV axis aligned along <111> direction with $\theta_{NV} \approx 54.7°$, $\varphi \approx 0°$. Meanwhile, both samples were fabricated into cross devices with a width of 10 μm and a length of 50 μm using micro-nanofabrication methods, and the Hall effect measurements were executed at room temperature with a DC current of 1 mA and a magnetic field applied along the film normal direction.

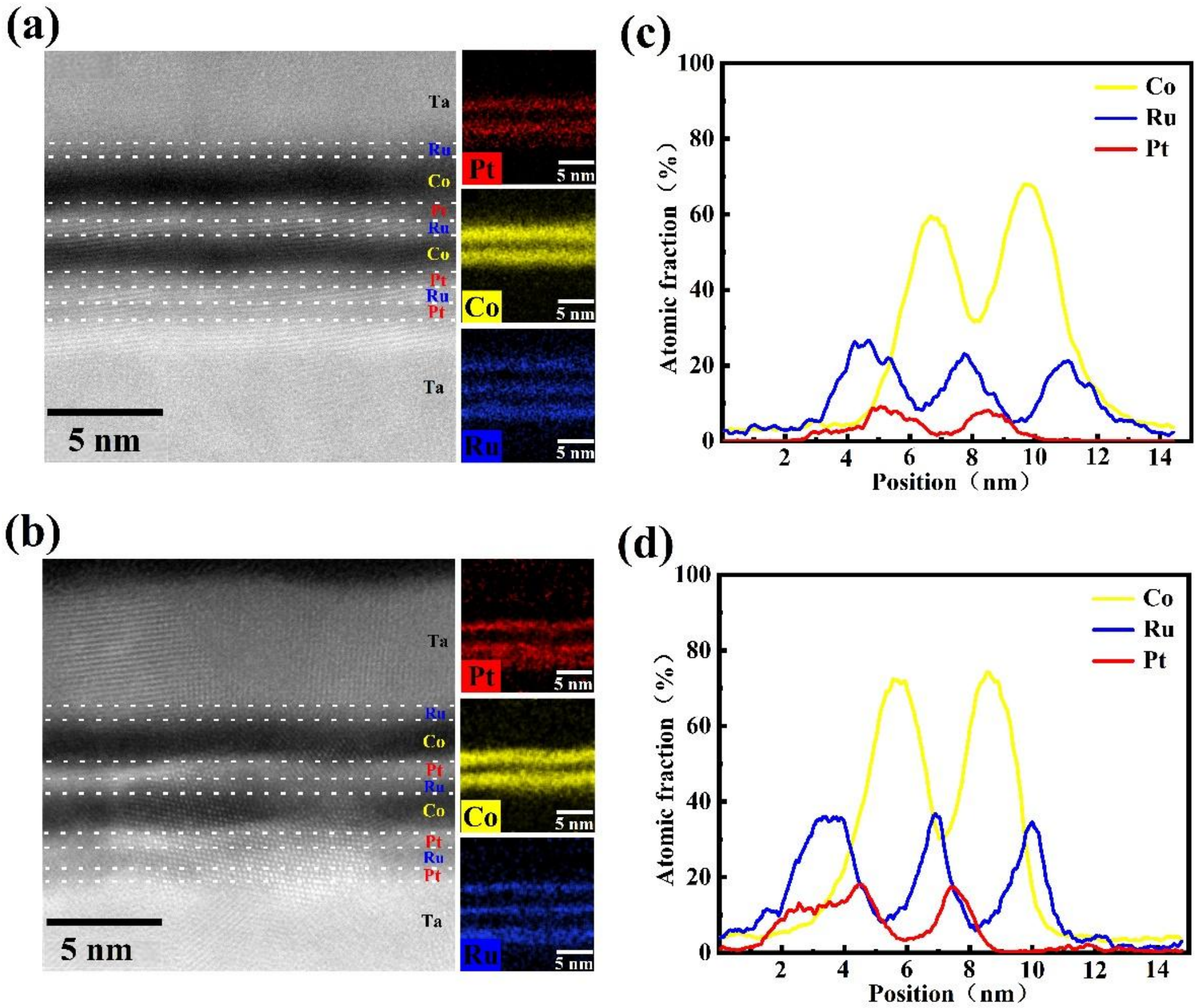

**FIG. 2.** Microstructure and elemental distribution of the SAF samples. (a), (b) Corresponding TEM images and elemental mapping of Pt, Co, and Ru for the non-compensated ($x_1$ = 1.6 nm, $x_2$ = 1.8 nm) and compensated ($x_1 = x_2$ = 1.7 nm) SAF structural variants, respectively. (c), (d) Line-scan profiles of the principal constituent elements (Pt, Co, Ru) across the two samples.

In order to characterize the structure of the samples, TEM measurements were executed. Figures 2(a) and (b) present the TEM images of the two samples with asymmetric ($x_1$ = 1.6 nm and $x_2$ = 1.8 nm) and symmetric ($x_1 = x_2$ = 1.7 nm) configurations, as well as the corresponding elemental resolution mapping images of Co, Pt, Ru. Besides, the line-scan profiles of the elements are shown in Figs. 2(c) and (d). The TEM results demonstrate atomic-level precision in thickness control across the multilayers and further reveal interfacial atomic diffusion at Pt/Co, Co/Ru, and Ru/Pt interfaces, manifested by the blurred boundaries of the layers in the TEM images and the overlap of the layers at the interfaces in the elemental mapping images and the line-scan profiles.

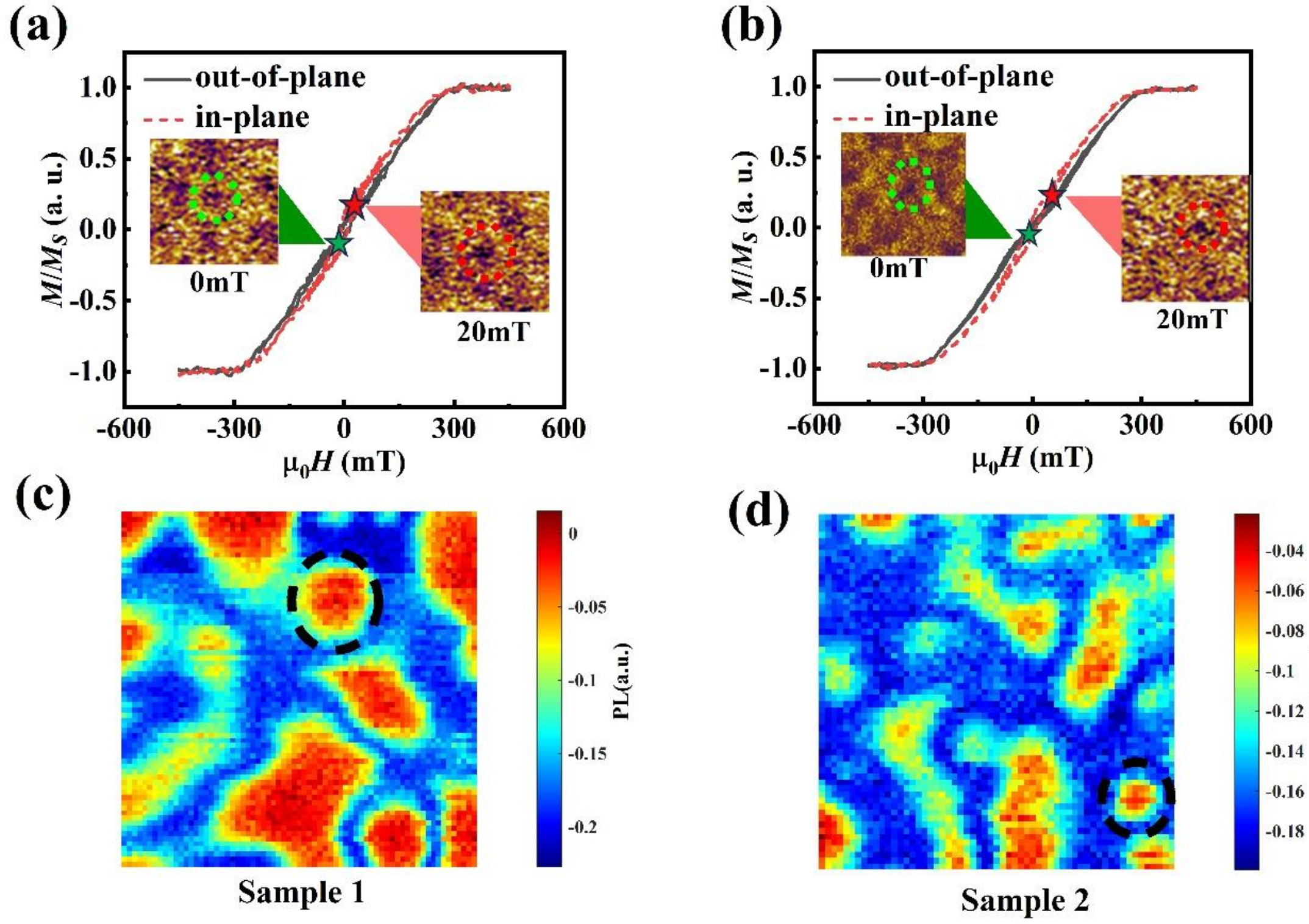

**FIG. 3.** Hysteresis loops and the domain structures of the SAF samples. (a) and (b) are the hysteresis loops of Sample 1 ($x_1$ = 1.6 nm, $x_2$ = 1.8 nm) and Sample 2 ($x_1$ = $x_2$ = 1.7 nm), respectively, as well as the MFM images of the SAF skyrmions in these two samples under zero field and non-zero field. The black solid lines and red dashed lines in (a) and (b) represent the out-of-plane and in-plane hysteresis loops, respectively. The insets marked by green and red star symbols show the MFM images of SAF skyrmions at $\mu_0 H$ = 0 and 20 mT, respectively. The magnetic fields are applied along the normal direction. The side lengths of the insets are 1 μm. (c) and (d) are the SNVM magnetometry image of Sample 1 and Sample 2, respectively. SAF skyrmions are marked by the dash line circles. The scale bar indicates the fluorescence intensity, and the side lengths of the insets are 1 μm.

To determine the appropriate thickness of the two Co layers for the existence of SAF skyrmions, specifically the thickness that brings the magnetic anisotropy to the critical state of transitioning from perpendicular to in-plane, VSM measurements were performed on the two series of samples. The detailed results of the out-of-plane[29] hysteresis loops are provided in Supplementary Materials Section 3. As shown in Fig.

S(3), for the non-compensated SAF series, the out-of-plane hysteresis loop exhibits the critical state for the sample with $x_1$ = 1.6 nm and $x_2$ = 1.8 nm (defined as Sample 1). For the compensated SAF series, the same critical state emerges in the loop of the sample with $x_1 = x_2$ = 1.7 nm (defined as Sample 2). Figs. 3(b) and (c) display the in-plane and out-of-plane hysteresis loops for Sample 1 and Sample 2. Notably, the saturation fields of the in-plane and out-of-plane loops are nearly identical, indicating that both structures are in a state of spin reorientation transition between in-plane and out-of-plane configurations, which is the state where SAF skyrmion states are likely to emerge.

To observe the SAF skyrmions in these two samples, MFM measurements under a magnetic field applied along the film normal direction are performed. The skyrmion images are shown in the insets of Figs. 3(a) and (b) (the detailed MFM images are shown in Supplementary Materials Section 4). The insets of Fig. 3(a) present the MFM images of a single SAF skyrmion in Sample 1 under zero field and an external field of 20 mT. Stable SAF skyrmions are observed in Sample 1 under zero field and a finite field[41]. These SAF skyrmions have a diameter of ~250 nm[42], and show no significant change under an external field of 20 mT. The insets of Fig. 3(b) present the MFM images of a single SAF skyrmion in Sample 2 under zero field and an external field of 20 mT. It can be preliminarily identified that the SAF skyrmions also exist in Sample 2 with a diameter of ~150 nm under zero field and show no significant change under an external field of 20 mT. Furthermore, the size of the SAF skyrmions in Sample 2 is noticeably smaller than that in Sample 1, which is consistent with the results in previous studies that the SAF skyrmions with zero net magnetic moment are smaller compared to the SAF skyrmions with non-zero net magnetic moment[29,39].

Conventional MFM faces challenges in imaging antiferromagnetic textures due to its reliance on detecting stray fields near the sample surface, which are typically weak in antiferromagnetic structures[43]. To overcome this limitation, we employed SNVM measurements, which offers significantly enhanced sensitivity to weak stray fields. Figs. 3(c) and (d) present the SNVM imaging results for Sample 1 and Sample 2 under zero field, respectively. The images demonstrate the coexistence of SAF stripe domains and skyrmion states in both samples, and the skyrmion sizes agree with previous MFM

observations.To investigate the electrical characteristics of SAF skyrmions, THE measurements were executed on both samples to quantify their topological Hall resistivity. According to previous studies, the topological Hall resistivity can be obtained by calculating the difference between the experimentally measured Hall resistivity ($\rho_{\mathrm{H}}$) and the fitted Hall resistivity ($\rho_{\mathrm{H}}^{\mathrm{fit}}$), where $\rho_{\mathrm{H}}^{\mathrm{fit}}$ includes contributions only from the normal Hall effect and the anomalous Hall effect. The field-dependent Hall resistivity ($\rho_{\mathrm{H}}$-$H$) was measured using standard anomalous Hall effect testing configurations (the device parameters detailed in Methods and Supplementary Section 1). The theoretical $\rho_{\mathrm{H}}^{\mathrm{fit}}$ was calculated using:

$$\rho_{\mathrm{H}}^{\mathrm{fit}}(H) = R_{\mathrm{o}}H + R_{\mathrm{s}}M(H) \qquad \text{Eq. (2)}$$

where $R_{\mathrm{o}}$, $R_{\mathrm{s}}$, and $M(H)$ are the ordinary Hall effect coefficient, the anomalous Hall effect coefficient, and the magnetization measured via VSM, respectively[43]. After obtaining $\rho_{\mathrm{H}}$ and $\rho_{\mathrm{H}}^{\mathrm{fit}}$, the topological Hall resistivity $\rho_{\mathrm{H}}^{\mathrm{T}}(H)$ can be determined using: $\rho_{\mathrm{H}}^{\mathrm{T}}(H) = \rho_{\mathrm{H}}(H) - \rho_{\mathrm{H}}^{\mathrm{fit}}(H)$. Figure 4(a) shows the $\rho_{\mathrm{H}}(H)$ and $\rho_{\mathrm{H}}^{\mathrm{fit}}(H)$ curves of Sample 1, with the magnetic field scanned from +400 mT to -400 mT. Significant deviation between these two curves is observed within $\pm$200 mT, corresponding to the field range where the SAF skyrmions were identified by the MFM measurements. The $\rho_{\mathrm{H}}^{\mathrm{T}}(H)$ curve for Sample 1 is obtained by calculating the difference between $\rho_{\mathrm{H}}(H)$ and $\rho_{\mathrm{H}}^{\mathrm{fit}}(H)$, as shown in Fig. 4(b). The $\rho_{\mathrm{H}}^{\mathrm{T}}(H)$ curve for Sample 1 exhibits two peaks of opposite signs between +200 mT and -200 mT, which are consistent with the THE of skyrmions. Therefore, we attribute these peaks to the THE of the non-compensated SAF skyrmions in Sample 1. The magnitudes of the two peaks are not equal, which is likely caused by the formation of magnetic domains induced by interfacial anisotropy, consistent with the expected characteristics of Néel-type skyrmions[43]. Additionally, we found that when $H = 0$, the value of $\rho_{\mathrm{H}}^{\mathrm{T}}$ is not zero, which also aligns with the MFM results showing the existence of zero-field skyrmions in Sample 1.

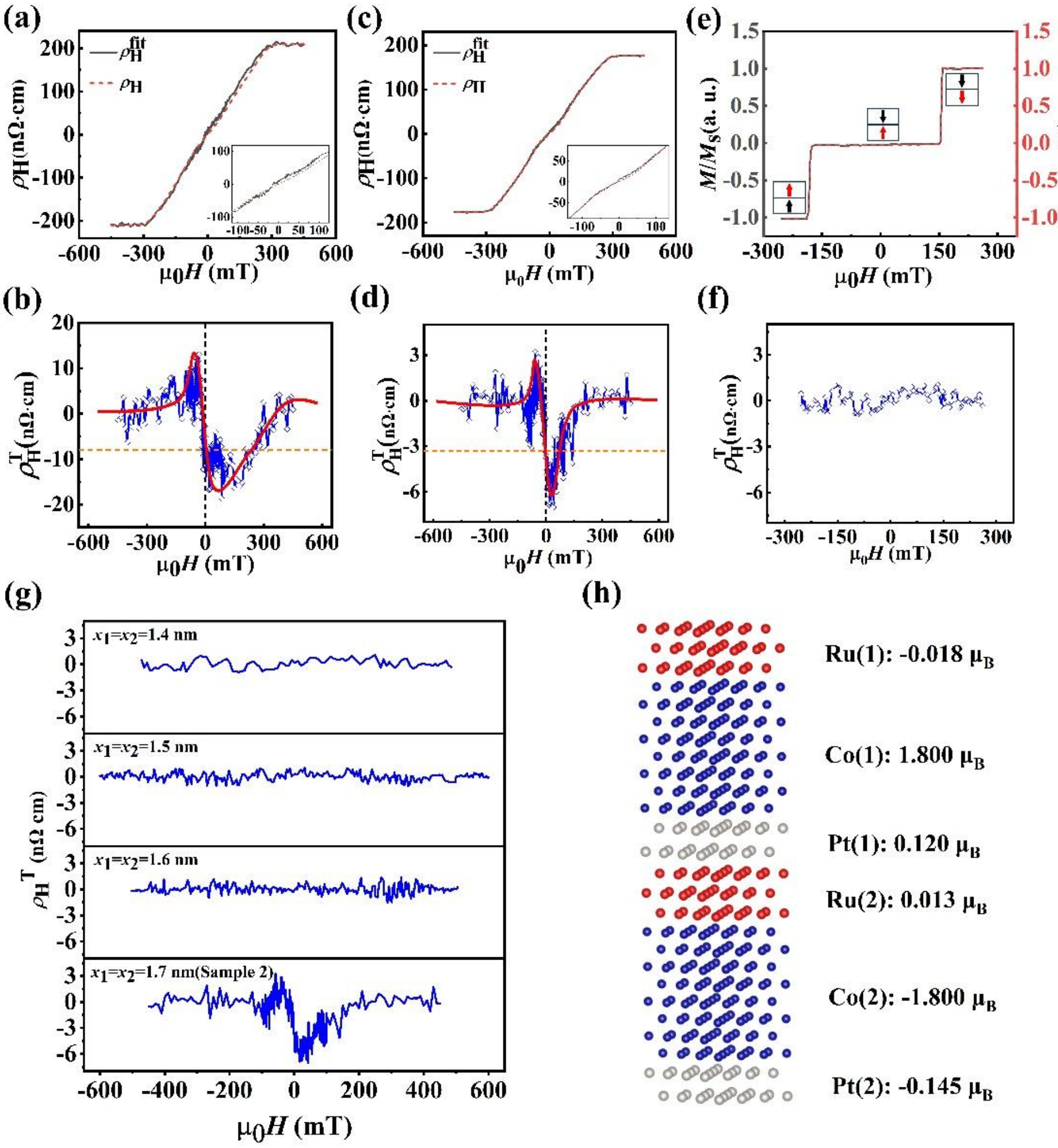

FIG. 4. Results of the THE measurements and the first-principles calculation for the SAF samples. (a) and (b) present the $\rho_{\mathrm{H}}$ and $\rho_{\mathrm{H}}^{\mathrm{T}}$ results for Sample 1, respectively. (c) and (d) display the $\rho_{\mathrm{H}}$ and $\rho_{\mathrm{H}}^{\mathrm{T}}$ results for Sample 2, respectively. In (a) and (c), the black solid lines and red dashed lines represent the $\rho_{\mathrm{H}}^{\mathrm{fit}}$ and $\rho_{\mathrm{H}}$. The insets in (a) and (c) show the details of the separation between the $\rho_{\mathrm{H}}^{\mathrm{fit}}$ and $\rho_{\mathrm{H}}$ curves for the two samples. In (b) and (d), the red solid lines serve as eye guides for the $\rho_{\mathrm{H}}^{\mathrm{T}}$ trend. (e) and (f) show the $\rho_{\mathrm{H}}$ and $\rho_{\mathrm{H}}^{\mathrm{T}}$ results for the control sample. The black solid lines and red dashed lines in (e) represent the normalized M-H curve and $\rho_{\mathrm{H}}$-H curve of the control sample, respectively. (g) $\rho_{\mathrm{H}}^{\mathrm{T}}$ results of the samples with $x_1 = x_2 = 1.4$, 1.5, 1.6, and 1.7. (h) First-

principles calculation model and layer-resolved atomic magnetic moments for the SAF structure. Positive and negative values are assigned to moments aligned parallel and antiparallel to the film normal, respectively.

Figures 4(c) and (d) present the $\rho_{\mathrm{H}}(H)$, $\rho_{\mathrm{H}}^{\mathrm{fit}}(H)$, and $\rho_{\mathrm{H}}^{\mathrm{T}}(H)$ curves for Sample 2. Similar to Sample 1, a clear separation between $\rho_{\mathrm{H}}(H)$ and $\rho_{\mathrm{H}}^{\mathrm{fit}}(H)$ is observed for Sample 2 within ±200 mT, as shown in Fig. 4(c). Meanwhile, the two peaks of opposite signs and asymmetric magnitudes within ±200 mT also exist in the $\rho_{\mathrm{H}}^{\mathrm{T}}(H)$ curve of Sample 2, which is exhibited in Fig. 4(d), although the peak values are significantly smaller than those of Sample 1. The magnetic field range of these peaks also conforms with the range of the SAF skyrmion existence observed by MFM. It can also be found that $\rho_{\mathrm{H}}^{\mathrm{T}}$ is non-zero at zero field in Sample 2, which agrees with the zero-field SAF skyrmion state of Sample 2 measured by MFM. These results suggest that compensated SAF skyrmions can also exhibit a non-zero THE.

To systematically evaluate potential artifacts arising from sample preparation or measurement conditions, we fabricated control samples with structure [buffer layer]/Pt/Co($x_1$ nm)/Ru/Pt/Co($x_2$ nm)/Ru/Ta ($x_1 = x_2 = 1.4$, 1.5, and 1.6) and performed identical VSM and Hall measurements. Figure 4(e) shows the normalized *M-H* curve (black solid line) and $\rho_{\mathrm{H}}$-*H* curve (red dashed line) of the control sample with $x_1 = x_2 = 1.4$ under field sweeps from +250 to -250 mT. The normalized *M-H* curve of the control sample exhibits three plateaus, corresponding to the three magnetization states of the two Co layers in the control sample: parallel-positive magnetized, antiparallel, and parallel-negative magnetized, as illustrated by the insets of Fig. 4(e). Significantly, Fig. 4(e) shows that the normalized *M-H* curve and $\rho_{\mathrm{H}}$-*H* curve almost completely overlap. Figure 4(f) shows the $\rho_{\mathrm{H}}^{\mathrm{T}}(H)$ curve of the control sample. It can be observed that $\rho_{\mathrm{H}}^{\mathrm{T}}$ of the control sample shows no peak characteristic of a skyrmion state, which is consistent with the absence of a skyrmion state in this sample. Figure 4(g) exhibits the $\rho_{\mathrm{H}}^{\mathrm{T}}(H)$ curves of all the control samples as well as Sample 2. It can be noticed that as $x_1$ and $x_2$ increase from 1.4 to 1.7, the peaks of the THE appear immediately when $x_1 = x_2 = 1.7$ (Sample 2), further demonstrating that the peaks of the THE are induced by the SAF skyrmion state, which only exists when the magnetic anisotropy is at the critical state of changing from perpendicular to in-plane anisotropy. The *M*-*H* and the $\rho_H$-$H$ curves of the samples with $x_1 = x_2 = 1.5$ and 1.6 are shown in Fig. S5.

Theoretical predictions suggest that compensated SAF skyrmions should not induce a THE, which disagrees with our experimental results. Compared to the $\rho_H^T$ of Sample 1 (~14.7 nΩ·cm), the $\rho_H^T$ of Sample 2 (~7.1 nΩ·cm) decreased by only about 50%. In contrast, there is a significant difference in the net magnetic moment between the two samples. Therefore, there must be additional factors contributing to the extra THE in the SAF samples other than the skyrmions in Co layers. The additional THE can be attributed to the contribution of the THE in the Pt and Ru layers. Due to the magnetic proximity effect, the Co layers can induce non-zero net magnetic moments in the adjacent Pt and Ru layers. Thus, the skyrmions in a single Co layer can also induce skyrmions in the Pt and Ru layers[44]. These induced skyrmions also contribute to the THE. Furthermore, the magnetic moments and texture of the Pt(Ru) layer between the two Co layers should be necessarily different from that of the Pt(Ru) layers outside the two Co layers. This asymmetry may result in non-zero net magnetic moments in the total system, thereby contributing to the nonzero skyrmion number and additional THE. Meanwhile, TEM analysis in Fig. 2 directly confirms the atomic diffusion at the Pt/Co, Co/Ru, and Pt/Ru interfaces. The interfacial atomic diffusion can enhance the magnetic moments at the interfaces of the Pt(Ru) layers induced by the magnetic proximity effect, thus enhancing the THE. Compared to the $\rho_H^T$ in room-temperature ferromagnetic skyrmion systems, THE in non-compensated and compensated SAF skyrmion systems is relatively weak but can still be experimentally observed. Therefore, the THE can serve as an electrical detection method for SAF skyrmions.

Furthermore, first-principles calculations were used to demonstrate the existence of the net magnetic moments in the compensated SAF skyrmion systems. The structure of the considered model is Pt/Co/Ru/Pt/Co/Ru heterostructure, which is similar to that of Sample 2, and shown in Fig. 4(h). The detailed information on the calculation model and the method is shown in Section 6 of the Supplementary Materials. By performing the calculation, the average atomic magnetic moments of each layer are obtained and are also shown in Fig. 4(h). The result demonstrates that Co(1) and Co(2) layers can induce net magnetic moments in adjacent Pt and Ru layers. The atomic magnetic moments of Co(1) and Co(2) layers show the same value and opposite orientation ($\pm 1.8\mu_B$). However, because of the broken inversion symmetry in this kind of multilayers, the atomic magnetic moments of Pt(1) ($0.120\mu_B$) and Pt(2) ($-0.145\mu_B$) show different values. The same condition also exists between the atomic magnetic moments of Ru(1) ($-0.018\mu_B$) and Ru(2) ($0.013\mu_B$). Thus, the heterostructure possesses a net atomic magnetic moment of $0.064\mu_B$. Moreover, the interfacial atomic diffusion can amplify the proximity-effect-induced magnetic moment asymmetry among the non-magnetic layers, thereby enhancing the net magnetic moments. The skyrmions constituted by these net atomic magnetic moments can be a significant origin of the additional THE.

In summary, we reported the observation of THE induced by the SAF skyrmions in $[Pt/Co/Ru]_2$ based SAF systems. The robust SAF skyrmions were observed in both non-compensated and compensated $[Pt/Co/Ru]_2$ systems. Simultaneously, the THE is observed in the external field range where the SAF skyrmions exist, even in the fully compensated condition. First-principles calculations and TEM images confirmed that the magnetic proximity effect and the interfacial atomic diffusion can induce net magnetic moments in the non-magnetic Pt and Ru layers, which can form skyrmions in the non-magnetic layers and cause the THE. This work presents an electrical detection method for SAF skyrmions and contributes to the realization of spintronics devices based on SAF skyrmions.

## Acknowledgements

The AFM and the magnetic - field - dependent MFM tests were provided by Sheng Lu, Yuanyuan Wu, and Prof. Zhipeng Hou from South China Normal University. The

zero - field MFM tests were performed using the AtomEdge Pro equipment provided by Truth Instruments. The SNVM tests were provided by Ciqtek. The TEM measurements were performed by Suzhou Institute of Nano-Tech and Nano-Bionics, Chinese Academy of Sciences. The Hall effect tests were provided by Songshan Lake Materials Laboratory. This work was supported by the National Natural Science Foundation of China (Grant Nos. 12241403, 12304136, T2394474, and 12404128), National Key Research and Development Program of China (Grant Nos. 2020YFA0309300, 2023YFF0719200 and 2024YFA1408801), “Chunhui” international cooperation Program (Grant No. HZKY20220018), Guangdong Basic and Applied Basic Research Foundation (Grant Nos. 2022A1515110863 and 2023A1515010837), Guangdong Science and Technology Project (Grant No. 2023QN10X275), and Project for Maiden Voyage of Guangzhou Basic and Applied Basic Research Scheme (Grant No. 2024A04J4186).

## Author Contributions

Xinbao Geng and Guanqi Li contributed equally to this work. Xinbao Geng and Guanqi Li performed the sample fabrication, as well as the VSM, MFM, and anomalous Hall effect measurements and data analysis with assistance from Wenjing Hu and Wenjing Zhong. Xinbao Geng and Junlin Wang carried out the micromagnetic modeling. Guanqi Li performed the first-principles calculations. Xiangyu Zheng. and Zhongxiang Zhang performed the nanofabrication of the devices. Zhendong Chen, Xiangyu Zheng, Junlin Wang, and Jing Wu designed the experiments and supervised the project. Xiaoming Xiong and Yongbing Xu contributed to the scientific discussions and provided significant revision suggestions. Xinbao Geng and Zhendong Chen wrote the manuscript with input from all authors.

# References

[1] A. Fert, V. Cros, and J. Sampaio, "Skyrmions on the track," Nat. Nanotech **8**(3), 152–156 (2013).

[2] R. Tomasello, E. Martinez, R. Zivieri, L. Torres, M. Carpentieri, and G. Finocchio, "A strategy for the design of skyrmion racetrack memories," Sci. Rep. **4**(1), 6784 (2014).

[3] G. Finocchio, F. Büttner, R. Tomasello, M. Carpentieri, and M. Kläui, "Magnetic skyrmions: From fundamental to applications," J. Phys. D: Appl. Phys. **49**(42), 423001 (2016).

[4] K.M. Song, J.-S. Jeong, B. Pan, X. Zhang, J. Xia, S. Cha, T.-E. Park, K. Kim, S. Finizio, J. Raabe, J. Chang, Y. Zhou, W. Zhao, W. Kang, H. Ju, and S. Woo, "Skyrmion-based artificial synapses for neuromorphic computing," Nat. Electron. **3**(3), 148–155 (2020).

[5] C.H. Marrows, and K. Zeissler, "Perspective on skyrmion spintronics," Appl. Phy. Lett. **119**(25), 250502 (2021).

[6] C.H. Marrows, J. Barker, T.A. Moore, and T. Moorsom, "Neuromorphic computing with spintronics," Npj Spintronics **2**(1), 12 (2024).

[7] S. Yang, Y. Zhao, K. Wu, Z. Chu, X. Xu, X. Li, J. Åkerman, and Y. Zhou, "Reversible conversion between skyrmions and skyrmioniums," Nat. Commun. **14**(1), 3406 (2023).

[8] A. Fert, N. Reyren, and V. Cros, "Magnetic skyrmions: advances in physics and potential applications," Nat. Rev. Mater. **2**(7), 17031 (2017).

[9] K. Everschor-Sitte, J. Masell, R.M. Reeve, and M. Kläui, "Perspective: Magnetic skyrmions—overview of recent progress in an active research field," J. Appl. Phys. **124**(24), 240901 (2018).

[10] J. Luo, J. Xia, X. Liang, Y. Zhou, L. Li, L. Gu, and G. Zhao, "Dynamics of skyrmions in synthetic ferrimagnetic materials," Phys. Rev. B **110**(21), 214409 (2024).

[11] J. Wang, M. Strungaru, S. Ruta, A. Meo, Y. Zhou, A. Deák, L. Szunyogh, P.-I. Gavriloaea, R. Moreno, O. Chubykalo-Fesenko, J. Wu, Y. Xu, R.F.L. Evans, and R.W. Chantrell, "Spontaneous creation and annihilation dynamics of magnetic skyrmions at elevated temperature," Phys. Rev. B **104**(5), 054420 (2021).

[12] H. Yin, X. Zheng, J. Wang, Y. Zhou, B. Kuerbanjiang, G. Li, X. Lu, K. Tong, Y. Wang, J. Wu, V.K. Lazarov, R.F.L. Evans, R.W. Chantrell, J. Cai, B. Liu, H. Meng, and Y. Xu, "Defect-correlated skyrmions and controllable generation in perpendicularly magnetized CoFeB ultrathin films," Appl. Phys. Lett. **119**(6), 062402 (2021).

[13] S. Woo, K. Litzius, B. Krüger, M.-Y. Im, L. Caretta, K. Richter, M. Mann, A. Krone, R.M. Reeve, M. Weigand, P. Agrawal, I. Lemesh, M.-A. Mawass, P. Fischer, M. Kläui, and G.S.D. Beach, "Observation of room-temperature magnetic skyrmions and their current-driven dynamics in ultrathin metallic ferromagnets," Nat. Mater. **15**(5), 501–506 (2016).

[14] R. Juge, S.-G. Je, D.D.S. Chaves, L.D. Buda-Prejbeanu, J. Peña-Garcia, J. Nath, I.M. Miron, K.G. Rana, L. Aballe, M. Foerster, F. Genuzio, T.O. Menteş, A. Locatelli, F. Maccherozzi, S.S. Dhesi, M. Belmeguenai, Y. Roussigné, S. Auffret, S. Pizzini, G. Gaudin, J. Vogel, and O. Boulle, "Current-driven skyrmion dynamics and drive-dependent skyrmion hall effect in an ultrathin film," Phys. Rev. Appl. **12**(4), 044007 (2019).

[15] C.E.A. Barker, C. Parton-Barr, C.H. Marrows, O. Kazakova, and C. Barton, "Skyrmion motion in a synthetic antiferromagnet driven by asymmetric spin wave emission," (2025).

[16] X.Z. Yu, N. Kanazawa, W.Z. Zhang, T. Nagai, T. Hara, K. Kimoto, Y. Matsui, Y. Onose, and Y. Tokura, "Skyrmion flow near room temperature in an ultralow current density," Nat. Commun. **3**(1), 988 (2012).

[17] W. Jiang, X. Zhang, G. Yu, W. Zhang, X. Wang, M. Benjamin Jungfleisch, J.E. Pearson, X. Cheng, O. Heinonen, K.L. Wang, Y. Zhou, A. Hoffmann, and S.G.E. te Velthuis, "Direct observation of the skyrmion hall effect," Nat. Phys. **13**(2), 162–169 (2017).
[18] K. Litzius, I. Lemesh, B. Krüger, P. Bassirian, L. Caretta, K. Richter, F. Büttner, K. Sato, O.A. Tretiakov, J. Förster, R.M. Reeve, M. Weigand, I. Bykova, H. Stoll, G. Schütz, G.S.D. Beach, and M. Kläui, "Skyrmion hall effect revealed by direct time-resolved X-ray microscopy," Nat. Phys. **13**(2), 170–175 (2017).
[19] K. Zeissler, S. Finizio, C. Barton, A.J. Huxtable, J. Massey, J. Raabe, A.V. Sadovnikov, S.A. Nikitov, R. Brearton, T. Hesjedal, G. Van Der Laan, M.C. Rosamond, E.H. Linfield, G. Burnell, and C.H. Marrows, "Diameter-independent skyrmion hall angle observed in chiral magnetic multilayers," Nat. Commun. **11**(1), 428 (2020).
[20] B.L. Brown, U.C. Täuber, and M. Pleimling, "Effect of the magnus force on skyrmion relaxation dynamics," Phys. Rev. B **97**(2), 020405 (2018).
[21] C. Moreau-Luchaire, C. Moutafis, N. Reyren, J. Sampaio, C.A.F. Vaz, N. Van Horne, K. Bouzehouane, K. Garcia, C. Deranlot, P. Warnicke, P. Wohlhüter, J.-M. George, M. Weigand, J. Raabe, V. Cros, and A. Fert, "Additive interfacial chiral interaction in multilayers for stabilization of small individual skyrmions at room temperature," Nat. Nanotech. **11**(5), 444–448 (2016).
[22] F. Büttner, I. Lemesh, and G.S.D. Beach, "Theory of isolated magnetic skyrmions: From fundamentals to room temperature applications," Sci. Rep. **8**(1), 4464 (2018).
[23] K. Wu, S. Yang, Y. Zhao, X. Liang, X. Xing, and Y. Zhou, "Tunable skyrmion–edge interaction in magnetic multilayers by interlayer exchange coupling," AIP Adv. **12**(5), 055210 (2022).
[24] Y. Zhao, S. Yang, J. Xia, X. Li, K. Wu, Y. Zhou, K. Fan, X. Zhang, and Y. Zhou, "Domain wall dynamics in ferromagnet/Ru/ferromagnet stacks with a wedged spacer," Appl. Phys. Lett. **119**(2), 022406 (2021).
[25] K. Wu, Y. Zhao, H. Hao, S. Yang, S. Li, Q. Liu, S. Zhang, X. Zhang, J. Åkerman, L. Xi, Y. Zhang, K. Cai, and Y. Zhou, "Topological transformation of synthetic ferromagnetic skyrmions: Thermal assisted switching of helicity by spin-orbit torque," Nat. Commun. **15**(1), 10463 (2024).
[26] J. Barker, and O.A. Tretiakov, "Static and dynamical properties of antiferromagnetic skyrmions in the presence of applied current and temperature," Phys. Rev. Lett. **116**(14), 147203 (2016).
[27] M. Belmeguenai, H. Bouloussa, Y. Roussigné, M.S. Gabor, T. Petrisor, C. Tiusan, H. Yang, A. Stashkevich, and S.M. Chérif, "Interface dzyaloshinskii-moriya interaction in the interlayer antiferromagnetic-exchange coupled pt/CoFeB/ru/CoFeB systems," Phys. Rev. B **96**(14), 144402 (2017).
[28] W. Jiang, G. Chen, K. Liu, J. Zang, S.G.E. Te Velthuis, and A. Hoffmann, "Skyrmions in magnetic multilayers," Phys. Rep. **704**, 1–49 (2017).
[29] W. Legrand, D. Maccariello, F. Ajejas, S. Collin, A. Vecchiola, K. Bouzehouane, N. Reyren, V. Cros, and A. Fert, "Room-temperature stabilization of antiferromagnetic skyrmions in synthetic antiferromagnets," Nat. Mater. **19**(1), 34–42 (2020).
[30] J. Sampaio, V. Cros, S. Rohart, A. Thiaville, and A. Fert, "Nucleation, stability and current-induced motion of isolated magnetic skyrmions in nanostructures," Nat. Nanotech. **8**(11), 839–844 (2013).
[31] S. Komineas, and N. Papanicolaou, "Traveling skyrmions in chiral antiferromagnets," SciPost Phys. **8**(6), 086 (2020).
[32] V.T. Pham, N. Sisodia, I. Di Manici, J. Urrestarazu-Larrañaga, K. Bairagi, J. Pelloux-Prayer, R.

Guedas, L.D. Buda-Prejbeanu, S. Auffret, A. Locatelli, T.O. Menteş, S. Pizzini, P. Kumar, A. Finco, V. Jacques, G. Gaudin, and O. Boulle, "Fast current-induced skyrmion motion in synthetic antiferromagnets," Sci. **384**(6693), 307–312 (2024).

[33] C. Jin, C. Song, J. Wang, and Q. Liu, "Dynamics of antiferromagnetic skyrmion driven by the spin Hall effect," Appl. Phys. Lett. **109**(18), 182404 (2016).

[34] P.E. Roy, "Method to suppress antiferromagnetic skyrmion deformation in high speed racetrack devices," J. Appl. Phys. **129**(19), 193902 (2021).

[35] A. Salimath, F. Zhuo, R. Tomasello, G. Finocchio, and A. Manchon, "Controlling the deformation of antiferromagnetic skyrmions in the high-velocity regime," Phys. Rev. B **101**(2), 024429 (2020).

[36] R. Tomasello, V. Puliafito, E. Martinez, A. Manchon, M. Ricci, M. Carpentieri, and G. Finocchio, "Performance of synthetic antiferromagnetic racetrack memory: domain wall versus skyrmion," J. Phys. D: Appl. Phys. **50**(32), 325302 (2017).

[37] H. Velkov, O. Gomonay, M. Beens, G. Schwiete, A. Brataas, J. Sinova, and R.A. Duine, "Phenomenology of current-induced skyrmion motion in antiferromagnets," New J. Phys. **18**(7), 075016 (2016).

[38] X. Zhang, Y. Zhou, and M. Ezawa, "Magnetic bilayer-skyrmions without skyrmion hall effect," Nat. Commun. **7**(1), 10293 (2016).

[39] R. Juge, N. Sisodia, J.U. Larrañaga, Q. Zhang, V.T. Pham, K.G. Rana, B. Sarpi, N. Mille, S. Stanescu, R. Belkhou, M.-A. Mawass, N. Novakovic-Marinkovic, F. Kronast, M. Weigand, J. Gräfe, S. Wintz, S. Finizio, J. Raabe, L. Aballe, M. Foerster, M. Belmeguenai, L.D. Buda-Prejbeanu, J. Pelloux-Prayer, J.M. Shaw, H.T. Nembach, L. Ranno, G. Gaudin, and O. Boulle, "Skyrmions in synthetic antiferromagnets and their nucleation via electrical current and ultra-fast laser illumination," Nat. Commun. **13**(1), 4807 (2022).

[40] Y. Zhao, S. Yang, K. Wu, S. Li, Y. Gao, H. Hao, X. Zhang, S. Wang, Q. Liu, S. Zhang, Z. Chu, J. Åkerman, and Y. Zhou, "Identifying and exploring synthetic antiferromagnetic skyrmions," Adv. Funct. Mater., (2023).

[41] M. Hervé, B. Dupé, R. Lopes, M. Böttcher, M.D. Martins, T. Balashov, L. Gerhard, J. Sinova, and W. Wulfhekel, "Stabilizing spin spirals and isolated skyrmions at low magnetic field exploiting vanishing magnetic anisotropy," Nat. Commun. **9**(1), 1015 (2018).

[42] T. Dohi, S. DuttaGupta, S. Fukami, and H. Ohno, "Formation and current-induced motion of synthetic antiferromagnetic skyrmion bubbles," Nat. Commun. **10**(1), 5153 (2019).

[43] A. Soumyanarayanan, M. Raju, A.L. Gonzalez Oyarce, A.K.C. Tan, M.-Y. Im, A.P. Petrović, P. Ho, K.H. Khoo, M. Tran, C.K. Gan, F. Ernult, and C. Panagopoulos, "Tunable room-temperature magnetic skyrmions in Ir/Fe/Co/Pt multilayers," Nat. Mater. **16**(9), 898–904 (2017).

[44] Q. Shao, Y. Liu, G. Yu, S.K. Kim, X. Che, C. Tang, Q.L. He, Y. Tserkovnyak, J. Shi, and K.L. Wang, "Topological hall effect at above room temperature in heterostructures composed of a magnetic insulator and a heavy metal," Nat. Electron. **2**(5), 182–186 (2019).